# Large-scale modulation in the superconducting properties of thin films due to domains in the SrTiO$_3$ substrate

Shai Wissberg and Beena Kalisky[*]

*Department of Physics and Institute of Nanotechnology and Advanced Materials, Bar-Ilan University, Ramat-Gan, Israel*

Scanning superconducting quantum interference device measurements reveal large-scale modulations of the superfluid density and the critical temperature in superconducting Nb, NbN, and underdoped YBa$_2$Cu$_3$O$_{7-\delta}$ films deposited on SrTiO$_3$ (STO). We show that these modulations are a result of the STO domains and domain walls, forming below the 105 K structural phase transition of STO. We found that the flow of normal current, measured above the superconducting transition, is also modulated over the same domain structure, suggesting a modified carrier density. In clean STO, domain walls remain mobile down to low temperatures. Modulated superconductivity over mobile channels offers the opportunity to locally control superconducting properties and better understand the relations between superconductivity and the local structure.

DOI: 10.1103/PhysRevB.95.144510

## I. INTRODUCTION

SrTiO$_3$ (STO) is a cubic perovskite at room temperature, with a strontium atom at every corner of the cube [1,2]. Six oxygen atoms, positioned at the center of each face, form the vertices of an octahedron, with a titanium atom at the center. At $T = 105$ K, neighboring oxygen octahedra rotate in opposite directions, turning the cubic unit cell into a tetragonal one with one elongated axis [labeled c and a in Fig. 1(a)]. The unit cells elongate in different directions to relieve local strain gradients. This leads to the formation of twin domains within the STO, each with different orthogonal orientations of the tetragonal unit cells. It is common to distinguish between the three possible domains by referring to the angle the domain boundaries, or walls, create with the cubic ⟨100⟩ direction [Fig. 1(a)]: 0°, 45°, and 90°.

In conducting interfaces with STO, such as LaAlO$_3$/SrTiO$_3$, the twin structure of STO strongly influences local normal state electronic properties [3–5] and weakly modulates the superfluid density in the superconducting state [5]. In $\delta$-doped STO [6–8], the twin structure influences the critical temperature [9], $T$c. In these material systems the current flows either in the STO itself, at an interface with another oxide, or in a doping layer of a few nanometers. The coupling between the STO and the two-dimensional (2D) conducting layer is excellent [10,11]. STO is also commonly used as a substrate for growing thin films [12–18], and the goal of this research was to find out whether the STO twin structure has an effect on the electronic properties of well-known thin films that use it as a substrate. These films can be thick (non-2D) or even polycrystalline.

In this paper we show that the normal and superconducting properties of films grown on STO are modulated over domains in the STO. We show that the electronic properties of superconducting films are modulated over domains in the STO. We used scanning superconducting quantum interference device (SQUID) microscopy to map the superfluid density, critical temperature, and normal current flow in thin films of Nb, NbN, and underdoped YBa$_2$Cu$_3$O$_{7-\delta}$ (YBCO) and show that the superfluid density and $T$c are lower on stripy features, which are most likely domain walls in the STO. Maps of current flow in the normal state (above $T$c) reveal that the conductivity is also modulated over the same stripe configuration.

## II. EXPERIMENTAL

We use scanning SQUID [19,20] susceptometer to record simultaneously the local static magnetism, current flow, and susceptibility. The SQUID gradiometer converts the flux passing through its pickup loop into a detectable electric signal, with a period of $\Phi_0 = hc/(2e)$. By mapping flux as a function of position, we capture the static magnetic topography. We also map the flow of current in the sample by mapping the flux generated by the current flow. In order to separate the static magnetism from the response to current, we apply an alternating current and use a lock-in amplifier. Additionally, we image the local susceptibility by using a one-turn coil (field coil) around the pickup loop to create a local magnetic field in the area of the pickup loop. This component of the signal is separated by driving an alternating current in the field coil at a different frequency. SQUID recording of susceptibility and current flow are illustrated in Fig. 1(c).

The samples measured in this paper are superconducting thin films, with surface area of 5 mm by 5 mm, grown on untreated [100] STO. The NbN and Nb polycrystalline samples were deposited at room temperature via direct current magnetron sputtering. Underdoped YBCO (sample 6 in Table I) was deposited by laser ablation under 30 mTorr N$_2$ flow at a temperature of 500 °C. Different superconducting materials were used to demonstrate that the effect we measure is general and not restricted to a specific superconducting material.

We identify superconductivity by measuring the susceptibility signal, which is diamagnetic near a superconductor. We quantify the diamagnetic response by tracking the susceptibility signal as we bring the sensor close to the sample. High above the surface the signal is zero, because our SQUID is a gradiometer and shows the difference between the flux detected by two pickup loops at the front and back of the sensor. Both loops sense the applied field, and the total is zero. When we bring the front of the gradiometer close to a superconductor, the front pickup loop detects a smaller flux reading, because the local field is eliminated by the superconductor. As we approach, the signal increases until the sensor comes in contact with the surface and the signal no longer changes. We describe the full susceptibility signal, $S_{total}$, as the difference between the susceptibility detected in contact and the value recorded far away (20–30 $\mu$m above the surface) [inset to Fig. 1(d)]. The sample's diamagnetic response to the locally applied field is directly related to the local superfluid density [19,20,23–25]. All samples measured (Table I) showed a diamagnetic response. By recording the diamagnetic response as a function of position at a constant distance from the sample, we follow small modulations in the superfluid density and study their characteristics as a function of location, temperature, and material. Susceptibility maps in this paper [Fig. 1(d)] show the flux detected by the pickup loop ($\Phi_0$) as a response to the current in the field coil (A), units of $\Phi_0/A$. In maps of static magnetic flux, units of $\Phi_0$, we image vortices [Fig. 1(e), vortices are pinned at 4.2 K]. The keyhole shape of each vortex comes from the convolution of the $z$ component of the magnetic field and the SQUID's point spread function.

---
[*]beena@biu.ac.il

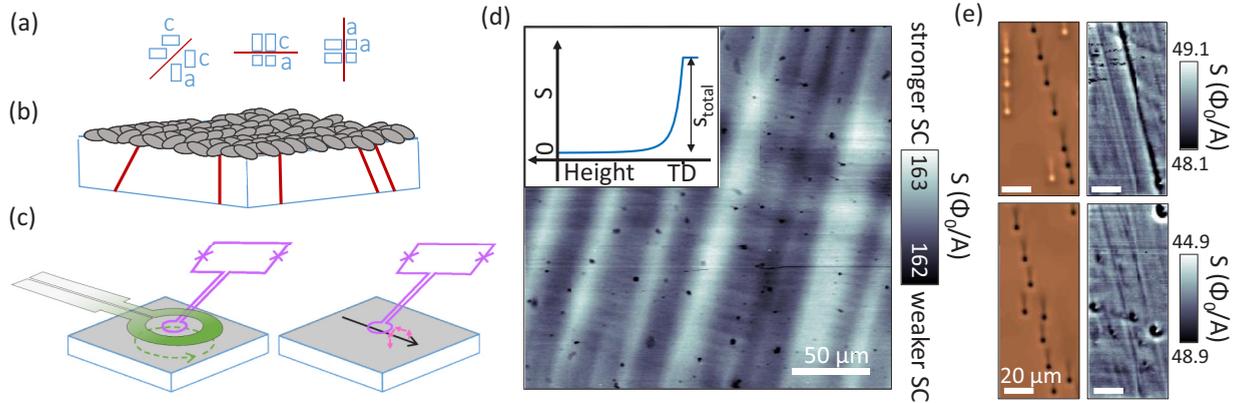

FIG. 1. Modulated superfluid density in superconducting films grown on STO. (a) Sketch of the three domain boundary orientations in STO after the structural phase transition. The 45° domain boundary (left) is between two domains with unit cells elongated in-plane, while the 0° (middle) and 90° (right) are due to lengthening out-of-plane. (b) A sketch of polycrystalline Nb on STO. Domain boundaries are represented by red lines in the STO. (c) Left: The way SQUID susceptometer detects the diamagnetic response of a superconducting sample. We induce current in the field coil (green loop), which generates the flow of supercurrents in the sample (green dashed line). The supercurrents counter the applied magnetic field. The pickup loop (purple) captures the overall field, which is less than the applied field. Right: SQUID magnetometer detects current distribution in the sample by mapping the magnetic fields generated by the current flow (black arrow). (d) Map of the diamagnetic response of sample 5 (Table I), showing stripy modulation of the superconductivity (SC) corresponding to the 0° and 90° domains of STO. Inset: A diagram of the susceptibility signal $S$ as a function of the distance between the SQUID sensor and the sample. The reading of the front minus the rear pickup loops increases as the sensor approaches the SC. After reaching the touchdown (TD) point, the susceptibility signal remains constant. The full susceptibility signal $S_{total}$ is the difference between the susceptibility signal at the TD point and far from the sample. (e) Magnetic (copper-colored) and susceptibility images, taken simultaneously, of sample 3 (Table I) at two different locations showing vortices aligned in lines corresponding to stripes of modulation in the diamagnetic response. Black and white objects are vortices of opposite polarity. Vortex density was controlled by an external magnetic field or by driving currents during cooldown. The stripy modulations are not the only factor determining the position of vortices, thus not all the vortices are aligned with the stripes. Each vortex carries one flux quantum, $\Phi_0$. The keyhole shape of the vortices is a result of a convolution between the magnetic field lines from the vortex (circular) and the point spread function of the SQUID (circle with leads).

## III. RESULTS

Maps of the diamagnetic response over the surface of the samples revealed, in all samples, stripes of modulated diamagnetic response [Fig. 1(d)]. These maps were recorded without applying transport current in the sample. The stripes were aligned along the $\langle 100 \rangle$ and $\langle 010 \rangle$ axes of the STO cubic crystalline directions. Darker stripes in our susceptibility maps indicate weaker diamagnetic response, implying lower superfluid density. The spacing between dark stripes ranges from too dense to resolve by our sensor up to 170 $\mu$m. This spacing is consistent with domain sizes mapped by optical polarized microscopy in STO samples [4,5,26]. The amplitude of these modulations ranges from 0.05% up to 2.86% of $S_{\text{total}}$. Maximal relative strength of modulations measured in each sample is summarized in Table I. In addition, the magnetic images of the sample [Fig. 1(e)] show that the vortices tend to pin on the stripes of reduced diamagnetic response, confirming that these stripes have smaller superfluid density.

TABLE I. Sample details. Superconducting transition $T$c was determined by transport measurements. The increase of the critical temperature in samples 2–5 agrees with previous studies [21,22].

| Sample no. | Material | Thickness (nm) | $T$c (K) | Scanned area (mm$^2$) | Maximal relative modulation measured (%) |
|---|---|---|---|---|---|
| 1 | NbN | 25 | 11.3 | 1.2 | 0.16 |
| 2 | Nb | 20 | 7.3 | 0.5 | 2.86 |
| 3 | Nb | 40 | 7.9 | 0.59 | 0.33 |
| 4 | Nb | 60 | 8.4 | 0.5 | 0.24 |
| 5 | Nb | 80 | 8.9 | 1.31 | 2.33 |
| 6 | Underdoped YBa$_2$Cu$_3$O$_{7-\delta}$ | 100 | 60 | 0.65 | 0.14 |

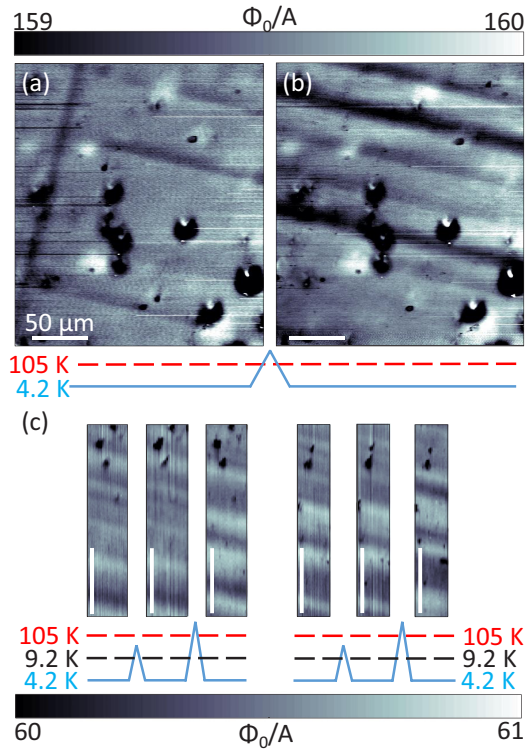

FIG. 2. New configuration of stripes with reduced superfluid density appears after recooling the sample through 105 K. (a) Susceptibility image showing stripy modulations in sample 3, corresponding to the 0° and 90° STO domains. The stripes of reduced diamagnetic response appear in an area of stronger signal. This suggests that the reduced signal occurs on domain walls. (b) The same area imaged after heating the sample to 140 K, above the structural transition temperature, and cooling it back to 4.2 K. The configuration of modulations changed, showing now only 0° domains. The black spots are defects of local reduction in the superfluid density, or bumps that elevate the SQUID and lower its signal. They can be used for comparing between the two images. Scale bar is 50 $\mu$m. (c) Two areas in sample 5 showing that the configuration of stripes did not change after cycling the temperature around 15 K, above the superconducting transition temperature. The stripe configuration changed after cycling around 140 K. Scale bars are 50 $\mu$m.

We attribute the modulation to STO domains and boundaries between them, based on their orientation, spacing, and behavior with temperature. Each time we cycled the temperature around the structural transition at 105 K, a new configuration of stripy modulations appeared (Fig. 2). Surface defects, black features in Figs. 2(a) and 2(b), do not change between cooldowns and enable comparing the same area in different cooldowns. When we cycled the temperature around $T$c, without crossing the temperature of the structural transition, the stripes appeared at the same locations [Fig. 2(c)]. In most of the data, the domains are too dense to tell whether the modulation is on the domain boundary or between different domains. In the occasions of spaced stripes, as observed in Figs. 2(a) and 2(b), thin stripes of lower susceptibility were observed. Limited by the size of our probe, we cannot differentiate between the scenario of wide bright domains coexisting with thin dark domains, and the scenario of bright domains separated by dark domain boundaries. However, in regions with spaced domain structure, areas of smaller superfluid density are consistently narrower than their surroundings (Fig. 2), supporting the scenario of reduced signal over domain boundaries.

Above $T$c, when the sample is no longer superconducting, the susceptibility map is not diamagnetic. We detect local $T$c by following the decay of the full susceptibility signal with increasing temperature. For example, in sample 5, Fig. 3(a) shows the temperature evolution of susceptibility taken on a wide stripe of brighter signal [black circle on the inset of Fig. 3(a)]. The diamagnetic response disappears above $T c_H = 8.5$ K, the highest value of the critical temperature measured over this stripe. Similar measurements in superconducting single crystals, just below $T$c, typically show that superconductivity nucleates in amorphous islands [23]. The superconducting islands grow as the temperature is lowered, until they merge and the entire sample is superconducting [23,27]. In our samples, superconductivity emerges in rectangular stripy regions that we identify as the STO domains. The inset of Fig. 3(a) shows that diamagnetism appears below $Tc_H$ in strong stripes. The modulation is also observed at 4.87 K.

The observation of modulated superconductivity in stripy regions that change location each time we cool down the sample through 105 K raises the question of whether $T$c is also modulated on these stripes. Figure 4(a) shows susceptibility along the cross section marked in the inset of Fig. 3. The line cut is taken across a stripe of stronger diamagnetic signal. The modulation of susceptibility across the stripe ($\Delta S$) increased with temperature and after reaching a critical value, decreased with the total susceptibility signal [Figs. 4(b) and 4(d)]. Close to $Tc_H$ the darker stripes are no longer superconducting. The relative part of the modulation from the full susceptibility signal ($\Delta S/S_{\text{total}}$) monotonously increased until $Tc_H$, where it reached 100% [Fig. 4(c)]. Comparing the temperature evolution of $S_{\text{total}}$ measured on and off the bright stripe [inset to Fig. 4(c)] shows that $T$c is lower by 1% (85 mK in this specific case). The maximal reduction of the local $T$c in different areas of this sample was 0.14 K. From these results, we conclude that the domain walls cause a reduction in the local $T$c.

Higher $T$c and stronger superfluid density in certain regions of the sample may come from higher electron density at these regions. We now turn to test this scenario by examining the flow of normal current in the sample above $T$c. We map the magnetic fields generated by the normal current by locking the SQUID magnetic reading to an alternating transport current in the sample, as illustrated in Fig. 1(c). We find that the map of magnetic flux representing the distribution of current flow in the sample also presents a stripy pattern [Fig. 5(a)]. Stripes of higher and lower flux response to current indicate that the current flow itself is modulated over channels [5,28]. Homogeneous current flow in a sample generates monotonously changing flux profiles. However, when the current density is modulated in the sample, the flux image obtains local changes of the signal [5,28]. This is what we observe in Figs. 5(a)–5(c). The modulation observed in Fig. 5(a) is small, but still detectable. Without knowing how much of the total current is included in the image, we cannot determine the extent of the modulation [29]. We estimate that it is small—a few percent at most.

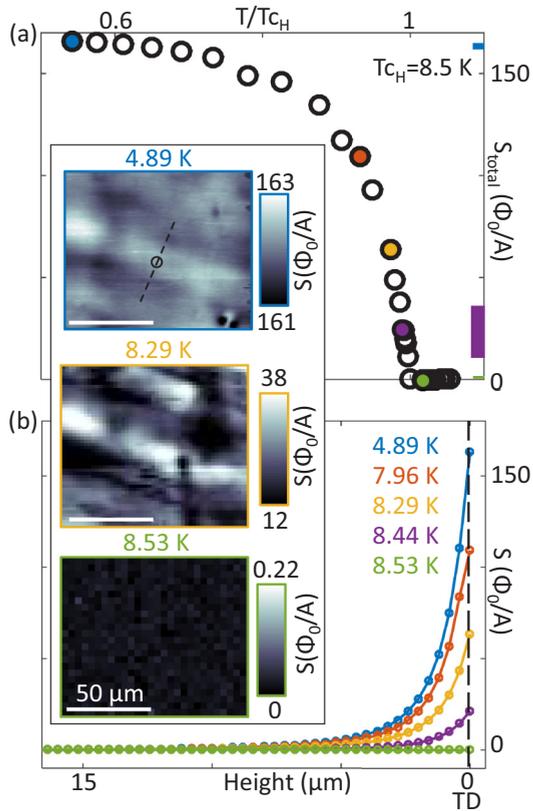

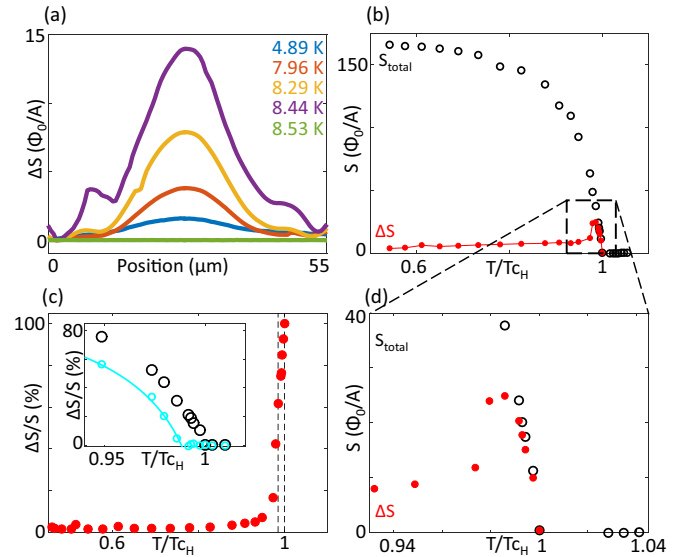

FIG. 3. Temperature evolution of the superfluid density. (a) The full susceptibility signal $S_{\text{total}}$ as a function of temperature. $Tc_H = 8.5$ K. Inset: Susceptibility maps of the same area in sample 4 (Table I) at three different temperatures corresponding to the colored points. The amplitude of the modulation increased with temperature, while above the critical temperature, the signal is not diamagnetic and does not depend on location. The dashed line marks the location of the cross section used in Fig. 4. Values span over 2, 26, and 0.22 $\Phi_0$/A (top to bottom). These ranges of values are also marked in panel a by a colored rectangle on the right-hand side. All scale bars are 50 $\mu$m. (b) Susceptibility as a function of height. As the sensor approaches the sample, the diamagnetic response becomes stronger (below $Tc$). The diamagnetic response near the surface is stronger at lower temperatures, because the penetration depth is smaller. $S_{\text{total}}$ is defined as the difference between the susceptibility signal at the TD point and the signal far from the sample. Above $Tc_H$, there is no difference between the two locations. Colors correspond to the colored points in panel a.

The amplitude of the modulation in the flux response to current over a stripe, $\Delta M$, depends on the temperature. Just below $Tc$, the current is flowing in a mixture of superconducting and normal regions—in our samples, stripy regions

FIG. 4. Modulation of the diamagnetic signal as a function of temperature. (a) Linecuts of susceptibility taken from the area indicated in Fig. 3 at different temperatures corresponding to the colors in Fig. 3. The modulation of the diamagnetic signal increased with temperature, disappearing above $Tc_H$. $\Delta S$ is defined as the amplitude of the modulation. (b) $S_{\text{total}}$ (black empty circles) plotted with $\Delta S$ (red circles) as a function of temperature in units of $T/Tc_H$. $\Delta S$ hardly changes with temperature until close to $Tc_H$, where it increases rapidly till it almost merges with $S_{\text{total}}$. (c) The relative modulation signal $\Delta S/S_{\text{total}}$ plotted as a function of temperature in units of $T/Tc_H$. The two extreme values of the critical temperature, the highest $Tc_H$ and the lowest $Tc_L$, are marked with dashed lines. Inset: $S_{\text{total}}$ signal as a function of temperature, measured on (cyan) and off (black) a stripe that showed reduced diamagnetic response. The darker stripes have lower $Tc$, $Tc_L$; $Tc_L/Tc_H = 0.99$. (d) Zoom on the region close to $Tc_H$ in panel b. $\Delta S$ increases and then decreases, overlapping $S_{\text{total}}$ only at $T/Tc_H = 1$.

(Fig. 3). Transport current flowing in the sample is expected to modulate strongly, avoiding the normal regions and preferring to flow in the superconducting regions. The susceptibility map in Fig. 5(b) shows stripy regions with a stronger diamagnetic response (white) compared with areas near them. As we decreased the temperature, the modulations in the superfluid density became smaller (Fig. 4). The current flow, which distributes between the channels according to the values of the local superfluid density, becomes less modulated as well [Fig. 5(c)]. The temperature dependence of the modulations in current flow [Fig. 5(d)] shows the increase of the amplitude until it peaked at $Tc$ and then decreased above $Tc$ to a nonzero value. Close to $Tc$, parts of the sample are still superconducting, while others are not. In this case, the modulations of the current are the strongest, since the current flows through the superconducting regions, avoiding the normal ones.

## IV. DISCUSSION

A stripy modulation of the superfluid density was observed a few years ago in single crystals of underdoped Ba(Fe$_{1-x}$Co$_x$)$_2$As$_2$ [30], which goes through structural transition from tetragonal to orthorhombic [31], where it forms twins. Enhanced superfluid density was detected on twin boundaries in Ba(Fe$_{1-x}$Co$_x$)$_2$As$_2$ and was related to the structural changes at the twin boundary [30]. Due to the higher superfluid density, vortices avoided pinning on the twin boundaries [32]. Similar behavior was observed in another pnictide, BaFe$_2$(As$_{1-x}$P$_x$)$_2$ [33]. The more typical behavior, observed in other twinned single-crystal superconductors (e.g., YBCO), is pinning of vortices on the boundaries [34–39]. In these materials, the twin boundaries are in the superconducting material itself, whereas in our paper, the twin planes are in the STO substrate. The twin planes cause a local reduction in the superfluid density of the superconducting film grown on top. We imaged vortex configurations simultaneously with the susceptibility measurements and observed a tendency of vortices to pin on the stripes of lower superfluid density [Fig. 1(e)].

There are several mechanisms that could explain the local change in superconducting properties due to domains in the substrate: (a) the dielectric constant is anisotropic and changes between domains [40]. (b) Polar domain walls could alter the local carrier density [41]. (c) Electrostatic potential modulates over domain walls [4], as well as the local current flow [5]. These modulations of properties over STO domains could lead to modulations in the superconductivity in systems that are strongly coupled to the STO, for example, LAO/STO [4,5] and $\delta$-doped STO [9]. (d) Formation of local contractions and strains caused by the STO domains. The domains bounded by the 0° and 90° walls are kinked at their boundaries, while 45° walls do not have kinks. These kinks could lightly bend the superconducting film along the domain wall. Strain is known to affect superconducting properties in global measurements of superconductors, specifically Nb [42–44]. The strains could be locally exerted at the nanoscale intersection between two adjacent domains. Although the Nb and NbN films were not epitaxially grown and are supposed to suppress strain effects in thick films, this explanation sits well with our observation of stripes in the 0° and 90° directions only. We cannot rule out the scenario that modulations of STO electronic properties induce modulations in the superconducting film, but we think it is less likely due to the grainy nature of our thin films.

Local reduction in the superfluid density could be a result of local changes in the pair potential or a difference in the local charge carriers [45]. Our measurements in the normal region show that the current modulates over the same STO domains. Although we cannot determine which of the mechanisms listed above is responsible for the modulation in superconducting properties, the modulation in the normal flow points to modulated carrier density as a favorable scenario.

STO domains remain mobile down to low temperature [46]. The high mobility of domains in clean STO leads to significant changes in domain configuration every cooldown through the structural transition temperature (Fig. 2). Domain walls in STO are also easily moved by electrostatic gating, even at low temperatures [4,5,26]. The electronic effects that are related to the domain structure move with the domain structure. Similarly, the modulated superfluid density that we observe changes dramatically between cooldowns and are expected to move with the domains. This may be exploited for controlling the value of local superfluid density (e.g., for creating SQUIDs). Such devices could be moved at low temperature and do not require extra fabrication steps.

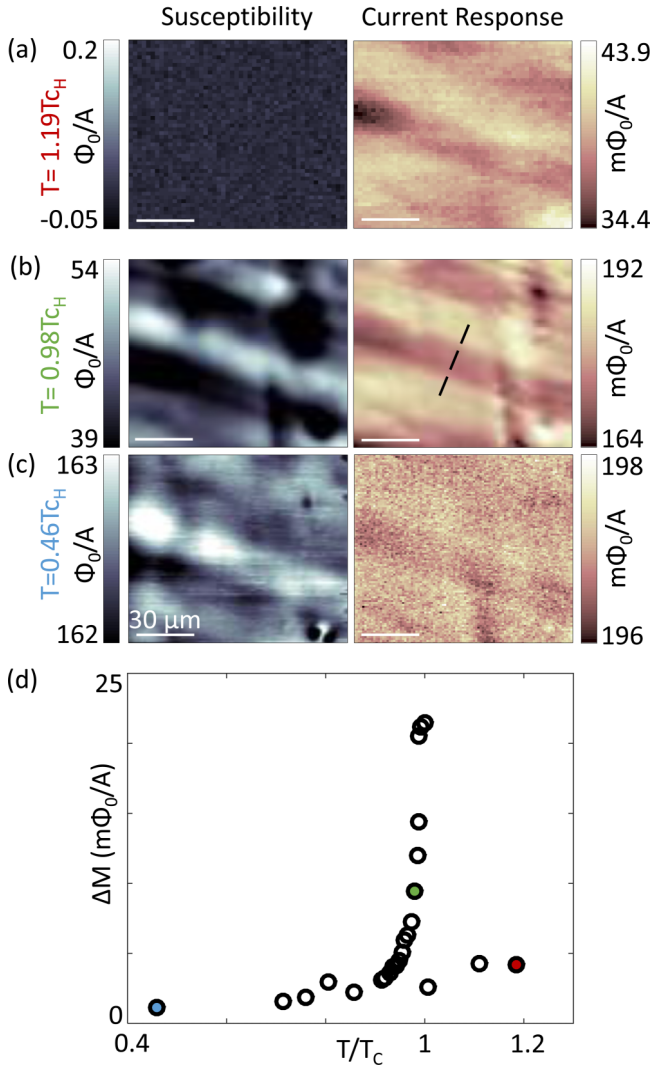

FIG. 5. The flow of normal current is also modulated, even above $Tc$. (a)–(c) Susceptibility (left, gray) and flux response to current (right) measured simultaneously on sample 4. (a) $T = 9.97$ K, above $Tc_H$. Stripy pattern is visible in the current response map, indicating that the normal current is modulated. The amplitude of the modulation, $\Delta M$, is 4.2 m$\Phi_0$/A. In contrast, no modulation, or any diamagnetic response, is detectable in the susceptibility map. (b) $T = 8.34$ K, below $Tc_H$. Stripy modulations are visible in both susceptibility and current response maps. The modulation observed in the current response, $\Delta M = 9.4$ m$\Phi_0$/A, is 224% stronger than the modulation observed above $Tc_H$. (c) $T = 4.22$ K. Weak stripy modulations in both susceptibility and current response. $\Delta M = 1.1$ m$\Phi_0$/A. (d) The modulation of the current response, $\Delta M$, as a function of temperature. $\Delta M$ is increased as we approach $Tc_H$ and drops immediately after to a nonzero value. Colored dots correspond to the images shown in panel (a)–(c). $\Delta M$ was extracted from the cross section marked in panel b.

**ACKNOWLEDGMENTS**

We thank A. Sharoni and G. Koren for providing the superconducting films and for helpful discussions. We thank

N. Vardi for help with the measurements. S.W. and B.K. were supported by the European Research Council Grant No. ERC-2014-STG-639792, Marie Curie Career Integration Grant No. FP7-PEOPLE-2012-CIG-333799, and Israel Science Foundation Grant No. ISF-1102/13.

[1] P. A. Fleury, J. F. Scott, and J. M. Worlock, Phys. Rev. Lett. **21**, 16 (1968).
[2] K. A. Müller, W. Berlinger, and F. Waldner, Phys. Rev. Lett. **21**, 814 (1968).
[3] J. F. Scott, E. K. H. Salje, and M. A. Carpenter, Phys. Rev. Lett. **109**, 187601 (2012).
[4] M. Honig, J. A. Sulpizio, J. Drori, A. Joshua, E. Zeldov, and S. Ilani, Nat. Mater. **12**, 1112 (2013).
[5] B. Kalisky, E. M. Spanton, H. Noad, J. R. Kirtley, K. C. Nowack, C. Bell, H. K. Sato, M. Hosoda, Y. Xie, Y. Hikita, C. Woltmann, G. Pfanzelt, R. Jany, C. Richter, H. Y. Hwang, J. Mannhart, and K. A. Moler, Nat. Mater. **12**, 1091 (2013).
[6] Y. Kozuka, Y. Hikita, C. Bell, and H. Y. Hwang, Appl. Phys. Lett. **97**, 012107 (2010).
[7] M. Kim, C. Bell, Y. Kozuka, M. Kurita, Y. Hikita, and H. Y. Hwang, Phys. Rev. Lett. **107**, 106801 (2011).
[8] M. Kim, Y. Kozuka, C. Bell, Y. Hikita, and H. Y. Hwang, Phys. Rev. B **86**, 085121 (2012).
[9] H. Noad, E. M. Spanton, K. C. Nowack, H. Inoue, M. Kim, T. A. Merz, C. Bell, Y. Hikita, R. Xu, W. Liu, A. Vailionis, H. Y. Hwang, and K. A. Moler, Phys. Rev. B **94**, 174516 (2016).
[10] A. Ohtomo and H. Y. Hwang, Nature **427**, 423 (2004).
[11] L. Li, C. Richter, J. Mannhart, and R. C. Ashoori, Nat. Phys. **7**, 762 (2011).
[12] W. Gong, J.-F. F. Li, X. Chu, Z. Gui, and L. Li, Appl. Phys. Lett. **85**, 3818 (2004).
[13] Y. Sun, W. Zhang, Y. Xing, F. Li, Y. Zhao, Z. Xia, L. Wang, X. Q. Ma, Q.-K. Xue, and J. Wang, Sci. Rep. **4**, 6040 (2014).
[14] S. Wang, L. Wang, and B. Gu, J. Mater. Sci. Technol. **24**, 899 (2008).
[15] H. N. Lee, A. Visinoiu, S. Senz, C. Harnagea, A. Pignolet, D. Hesse, and U. Gösele, J. Appl. Phys. **88**, 6658 (2000).
[16] P. Mele, K. Matsumoto, T. Horide, O. Miura, A. Ichinose, M. Mukaida, Y. Yoshida, and S. Horii, Supercond. Sci. Technol. **19**, 44 (2006).
[17] G. Koren, E. Polturak, B. Fisher, D. Cohen, and G. Kimel, Appl. Phys. Lett. **53**, 2330 (1988).
[18] B. Kalisky, P. Aronov, G. Koren, A. Shaulov, Y. Yeshurun, and R. P. Huebener, Phys. Rev. Lett. **97**, 067003 (2006).
[19] B. W. Gardner, J. C. Wynn, P. G. Björnsson, E. W. J. Straver, K. A. Moler, J. R. Kirtley, and M. B. Ketchen, Rev. Sci. Instrum. **72**, 2361 (2001).
[20] M. E. Huber, N. C. Koshnick, H. Bluhm, L. J. Archuleta, T. Azua, P. G. Björnsson, B. W. Gardner, S. T. Halloran, E. A. Lucero, K. A. Moler, P. G. Björnsson, B. W. Gardner, S. T. Halloran, E. A. Lucero, and K. A. Moler, Rev. Sci. Instrum. **79**, 053704 (2008).
[21] J. Kodama, M. Itoh, and H. Hirai, J. Appl. Phys. **54**, 4050 (1983).
[22] S. A. Wolf, J. J. Kennedy, and M. Nisenoff, J. Vac. Sci. Technol. **13**, 145 (1976).
[23] C. W. Hicks, T. M. Lippman, M. E. Huber, J. G. Analytis, J.-H. H. Chu, A. S. Erickson, I. R. Fisher, and K. A. Moler, Phys. Rev. Lett. **103**, 127003 (2009).
[24] J. R. Kirtley, C. C. Tsuei, K. A. Moler, V. G. Kogan, J. R. Clem, and A. J. Turberfield, Appl. Phys. Lett. **74**, 4011 (1999).
[25] V. G. Kogan, Phys. Rev. B **68**, 104511 (2003).
[26] Z. Erlich, Y. Frenkel, J. Drori, Y. Shperber, C. Bell, H. K. Sato, M. Hosoda, Y. Xie, Y. Hikita, H. Y. Hwang, and B. Kalisky, J. Supercond. Nov. Magn. **28**, 1017 (2014).
[27] J. A. Bert, K. C. Nowack, B. Kalisky, H. Noad, J. R. Kirtley, C. Bell, H. K. Sato, M. Hosoda, Y. Hikita, H. Y. Hwang, and K. A. Moler, Phys. Rev. B: Condens. Matter Mater. Phys. **86**, 60503 (2012).
[28] Y. Frenkel, N. Haham, Y. Shperber, C. Bell, Y. Xie, Z. Chen, Y. Hikita, H. Y. Hwang, and B. Kalisky, ACS Appl. Mater. Interfaces **8**, 12514 (2016).
[29] B. J. Roth, N. G. Sepulveda, and J. P. Wikswo, J. Appl. Phys. **65**, 361 (1989).
[30] B. Kalisky, J. R. Kirtley, J. G. Analytis, J.-H. H. Chu, A. Vailionis, I. R. Fisher, and K. A. Moler, Phys. Rev. B **81**, 184513 (2010).
[31] J.-H. H. Chu, J. G. Analytis, C. Kucharczyk, and I. R. Fisher, Phys. Rev. B **79**, 014506 (2009).
[32] B. Kalisky, J. R. Kirtley, J. G. Analytis, J.-H. Chu, I. R. Fisher, and K. A. Moler, Phys. Rev. B **83**, 064511 (2011).
[33] A. Yagil, Y. Lamhot, A. Almoalem, S. Kasahara, T. Watashige, T. Shibauchi, Y. Matsuda, and O. M. Auslaender, Phys. Rev. B **94**, 064510 (2016).
[34] L. Y. Y. Vinnikov, L. A. A. Gurevich, G. A. A. Yemelchenko, Y. A. A. Ossipyan, I. V. V. Grigoryeva, A. E. E. Koshelev, and Y. A. A. Osip'yan, Solid State Commun. **67**, 421 (1988).
[35] L. Y. Y. Vinnikov, L. A. A. Gurevich, I. V. V. Grigoryeva, A. E. E. Koshelev, and Y. A. A. Osip'yan, J. Less-Common Met. **164–165**, 1271 (1990).
[36] G. J. Dolan, G. V. Chandrashekhar, T. R. Dinger, C. Feild, and F. Holtzberg, Phys. Rev. Lett. **62**, 827 (1989).
[37] P. L. Gammel, C. A. Duran, D. J. Bishop, V. G. Kogan, M. Ledvij, A. Y. Simonov, J. P. Rice, and D. M. Ginsberg, Phys. Rev. Lett. **69**, 3808 (1992).
[38] J. A. Herbsommer, G. Nieva, and J. Luzuriaga, Phys. Rev. B **62**, 3534 (2000).
[39] V. K. Vlasko-Vlasov, L. A. Dorosinskii, A. A. Polyanskii, V. I. Nikitenko, U. Welp, B. W. Veal, and G. W. Crabtree, Phys. Rev. Lett. **72**, 3246 (1994).
[40] T. Sakudo and H. Unoki, Phys. Rev. Lett. **26**, 851 (1971).
[41] E. K. H. Salje, O. Aktas, M. A. Carpenter, V. V. Laguta, and J. F. Scott, Phys. Rev. Lett. **111**, 247603 (2013).
[42] D. C. Hill and R. M. Rose, Metall. Trans. **2**, 585 (1971).
[43] V. S. Bobrov and M. A. Lebyodkin, Mater. Sci. Eng. A **164**, 449 (1993).
[44] G. Pristáš, S. Gabánia, E. Gažoa, V. Komanickýb, M. Orendáčb, and H. Youc, Thin Solid Films **556**, 470 (2014).
[45] B. Lorenz and C. W. Chu, in *Frontiers in Superconducting Materials*, edited by A. V. Narlikar (Springer-Verlag, Heidelberg, 2005), pp. 459–497.
[46] A. V. Kityk, W. Schranz, P. Sondergeld, D. Havlik, E. K. H. Salje, and J. F. Scott, Phys. Rev. B **61**, 946 (2000).